\documentclass{iaus}
\usepackage{graphicx}

\title[The Galactic halo using beryllium as a time scale] 
{A view of the Galactic halo using beryllium as a time scale}

\author[Smiljanic et al.]   
{Rodolfo Smiljanic$^1$
 \thanks{Present address: ESO, Garching bei M\"unchen, Germany.},
  L. Pasquini$^2$,
  P. Bonifacio$^{3,4,5}$,
  D. Galli$^6$, \\
  B. Barbuy$^1$,
  R. Gratton$^7$,
 \and S. Randich$^6$
 }

\affiliation{$^1$IAG, Universidade de S\~ao Paulo, S\~ao Paulo, Brazil 
 \\ email: {\tt rodolfo@astro.iag.usp.br} \\[\affilskip]
$^2$ESO, Garching bei M\"unchen, Germany, 
$^3$GEPI Observatoire de Paris - Meudon, France, 
$^4$INAF, Osservatorio di Trieste, Trieste, Italy, 
$^5$CIFIST Marie Curie Excellence Team, 
$^6$INAF- Osservatorio di Arcetri, Firenze, Italy
$^7$INAF-Osservatorio di Padova, Padova, Italy, 
}

\pubyear{2009}
\volume{265}  
\pagerange{1--2}
\jname{Chemical abundances in the Universe: connecting first stars to planets}
\editors{K. Cunha, M. Spite \& B. Barbuy, eds.}
\begin{document}

\maketitle

\begin{abstract}
Beryllium stellar abundances were suggested to be a good tracer of time 
in the early Galaxy. In an investigation of its use as a cosmochronometer, 
using a large sample of local halo and thick-disk dwarfs, evidence was found that in 
a log(Be/H) vs.\, [$\alpha$/Fe] diagram the halo stars 
separate into two components. One is consistent with predictions of evolutionary 
models while the other is chemically indistinguishable from the thick-disk stars. 
This is interpreted as a difference in the star formation history of the two components and 
suggests that the local halo is not a single uniform population where a clear 
age-metallicity relation can be defined.
\end{abstract}

\firstsection 
\section{Introduction}

Be is a pure product of cosmic-ray spallation in the ISM involving mostly CNO nuclei. 
Abundances of Be in metal-poor stars show a linear relation with Fe (and O) with 
a slope close to one, implying that Be behaves as a primary element (\cite[Smiljanic et al. 2009, and references therein]{Sm09}). 
If cosmic rays are globally transported across the Galaxy, the production 
of Be is a widespread process and Be abundances should 
have a smaller scatter than the products of stellar nucleosynthesis at a given time in the early Galaxy 
(\cite[Suzuki \& Yoshii 2001]{SY01}). In other words Be should be a good tracer of time.

\cite[Pasquini et al. (2004, 2007)]{Pas04,07} calculated Be abundances in turn-off stars 
of two globular clusters, NGC 6397 and NGC 6752. The Be ages derived 
from a model of the evolution of Be with time (\cite[Valle et al. 2002]{Valle02}) were shown 
to agree with those derived from theoretical isochrones, supporting the use of Be as a 
cosmochronometer. 

\cite[Pasquini et al. (2005)]{Pas05} analyzed a sample of 20 
halo and thick-disk stars and found a possible separation between stars of the two components in a 
log(Be/H) vs.\, [$\alpha$/Fe] diagram.  This was interpreted as a difference in the time 
scales of star formation.

\cite[Smiljanic et al. (2009)]{Sm09} analyzed the largest sample of halo and 
thick-disk stars to date, extending the investigation of Be as cosmochronometer and its role 
as a discriminator of different stellar populations in the Galaxy. The thick disk was found to 
be a homogeneous population. The halo stars were found  to divide into 
two different components (Fig.\,\ref{fig1}).

\section{The Galactic halo}

\begin{figure}[t]
\begin{center}
 \includegraphics[width=4.6in]{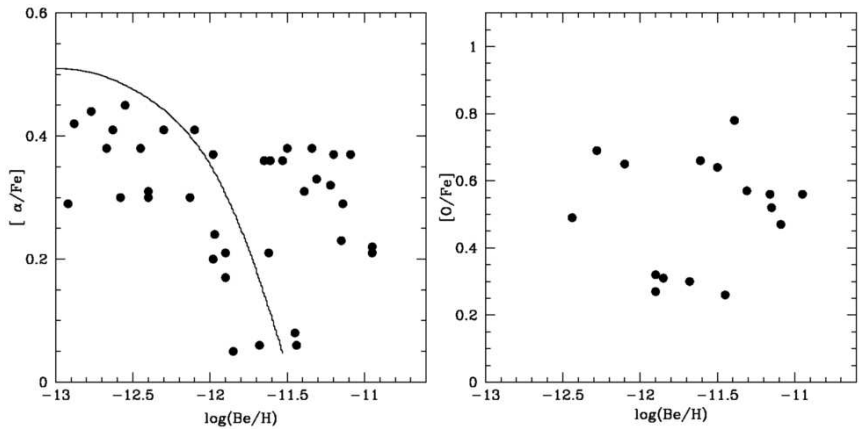} 
 \caption{(a) log(Be/H) vs.\, [$\alpha$/Fe] for the halo stars of \cite[Smiljanic et al. (2009)]{Sm09}. 
The solid line is the model prediction of \cite[Valle et al. (2002)]{Valle02}. (b) log(Be/H) vs.\, [O/Fe] 
with new preliminary oxygen abundances for 
a subsample of the halo stars., corrected for NLTE (\cite[Fabbian et al. 2009]{Fabbian09})}
   \label{fig1}
\end{center}
\end{figure}

In a diagram of [O/Fe] vs. log(Be/H) the abscissa can be considered as increasing 
time and the ordinate as the star formation rate. In \cite[Smiljanic et al. (2009)]{Sm09} 
oxygen abundances were not available, so mean abundances of $\alpha$-elements 
were used instead. Here we present new preliminary oxygen abundances determined 
from the infrared triplet at 777nm. 

As shown in Fig.\,\ref{fig1}, using either $\alpha$ or oxygen abundances, the halo stars define 
two clear distinct sequences. One sequence is chemically similar to the thick disk, the other 
agrees with the models of \cite[Valle et al. (2002)]{Valle02}. The latter, however, with 
[$\alpha$/Fe] $\leq$ 0.25 and log(Be/H) $\leq$ $-$11.4, have similar kinematics. The stars have mostly  
velocity in the direction of the rotation of the Galaxy V $\sim$ 0 and the perigalactic 
distance Rmin $\leq$ 1 kpc (\cite[see Smiljanic et al. 2009 for details)]{Sm09}, as expected 
for accreted stars. 

The splitting into two components may be related to the accretion of external systems or to variations of star 
formation in different and initially independent regions of the early halo. The interpretetion is still open, 
it is however clear that the halo is not a single uniform population with a single age-metallicity relation. 
A similar division was found by \cite[Nissen \& Schuster (1997, 2009)]{NS97,NS09} but using Fe as 
a tracer of time. The division is clearer when Be is used as a time scale. In the same line, recent simulations of the formation of disk 
galaxies in a $\Lambda$CDM universe by \cite[Zolotov et al. (2009)]{Zolotov09} show that the inner halo has a 
dual nature, it is composed both by `in situ stars' formed in the potential well of the galaxy and by 
accreted stars, formed in subhalos later accreted by the galaxy.

\begin{acknowledgments}
R.S. acknowledges financial support from FAPESP (04/13667-4 and 08/55923-8).
\end{acknowledgments}

\end{document}